%% file: 00-main.tex
\documentclass[letterpaper]{article} 
\usepackage{aaai2027}  
\usepackage[hyphens]{url}  
\usepackage{graphicx} 
\usepackage{natbib}  
\usepackage{caption} 
\usepackage{algorithm}
\usepackage{algpseudocode}

\usepackage{algorithmicx}
\usepackage{amsmath}
\algnewcommand{\LeftComment}[1]{\Statex \(\triangleright\) #1}
\usepackage{threeparttable}
\usepackage{algpseudocode}
\usepackage{booktabs}
\usepackage{multirow}
\usepackage{rotating}
\newcommand{\se}[1]{\ensuremath{{\scriptstyle\,\pm\,#1}}}

\usepackage{amsmath}
\usepackage{listings}
\DeclareCaptionStyle{ruled}{labelfont=normalfont,labelsep=colon,strut=off} 
\floatstyle{ruled}
\newfloat{listing}{tb}{lst}{}
\floatname{listing}{Listing}

\usepackage{booktabs}

\title{Evolve on the Host, Predict on the Edge: Deploying Online Neuroevolutionary Architecture Search for Cross-sectional Stock Return Prediction}
\author{
    Jonathan Chang\textsuperscript{\rm 1},
    Zimeng Lyu\textsuperscript{\rm 2}
}
\affiliations{
    \textsuperscript{\rm 1} Union County Magnet High School \\
    \textsuperscript{\rm 2}Department of Computer Science and Technology, Kean University\\
    jonathanbokeunchang@gmail.com, zimeng.lyu@kean.edu
}

\begin{document}

\maketitle

\begin{abstract}
Accurate forecasting models are usually large, expensive to update
online, and fixed in architecture once trained. We apply ONE-NAS, an
online neuroevolutionary architecture search that evolves a population
of small recurrent networks as each window of data arrives, to daily
cross-sectional stock return prediction, and pilot it on a host and
endpoint pipeline: the host runs the search and ships each generation's
champion genomes over TCP/IP to a Raspberry Pi 4B, which predicts
online. On the Pi a single champion predicts a 50-stock window in
24.6~ms and the ensemble of 40 island champions in 556~ms, far inside
the daily decision cycle. On four panels of US mid-cap equities over
2022--2024, reading the population as a rank-mean ensemble of island
champions returns $+27.5\%$ net of realised transaction costs, against
$+11.3$ to $+14.8\%$ for online LSTM, online GRU and monthly-retrained
LSTM baselines and $+4.5\%$ for the single best genome used in prior
ONE-NAS work. 
\end{abstract}

\input{01-Introduction}
\input{02-RelatedWork}
\input{03-Method}

\input{04-Dataset}
\input{05-Results}
\input{06-Conclusion}

\bigskip


\bibliography{99-References}

\clearpage
\appendix
\input{80-Appendix}

\input{81-AppendixResults}


\end{document}

%% file: 01-Introduction.tex
\section{Introduction}

Real-world time series forecasting tasks usually face data drift: a model fitted once on historical data degrades as the generating process drifts from its training sample. Online forecasting updates the model as new observations arrive \cite{pham2022learning,wen2023onenet,zhao2025proactive}, but accurate forecasting models are usually large \cite{das2023decoder,ansari2024chronos,woo2024unified}: expensive to update at the cadence of the stream, expensive to serve where the data are produced, and fixed in architecture once trained, so only their weights can adapt.

The problem is acute in stock forecasting, where daily returns have a low signal-to-noise ratio and the feature-to-return relationship shifts with market regimes, so a model fitted to one period often fails in the next \cite{zhao2023doubleadapt}. Deep stock models \cite{li2024master} are refitted on a schedule, and incremental learning \cite{zhao2023doubleadapt} updates their weights between refits, but in both cases the architecture is fixed in advance.

We address adaptation with online neural architecture search. ONE-NAS \cite{lyu2023onenas} extends evolutionary architecture search \cite{liu2021survey} to the online setting: a population of recurrent neural networks, organised in islands, evolves as each window of data arrives, with offspring trained briefly from inherited parent weights and survivors selected on the most recent data. Architectures and weights both adapt, so there is no offline training phase to go stale. We apply ONE-NAS to daily cross-sectional stock return forecasting on CRSP (Center for Research in Security Prices) data, and show that the result depends on how the population is read: the single best genome of prior ONE-NAS work trails every neural baseline, while a rank-mean ensemble of the island champions outperforms them all.

An online neuroevolutionary NAS algorithm deploys as two workloads: the parallel and expensive search runs on a cloud service or a workstation, and the forecasting, cheap because the evolved networks are small, runs on an edge endpoint device (Figure~\ref{fig:pipeline}). Each generation the host sends the champion genomes over TCP/IP to a Raspberry Pi, which forecasts as each day's features arrive and keeps forecasting from the last received genomes if the host is unavailable. The Pi is a  constrained target: inference that meets the daily cadence there does so on any hardware a practitioner owns.

The contributions of this paper are:

\begin{itemize}
\item A deployment workflow for stock forecasting and for trading on the forecasts, with ONE-NAS running the online neuroevolutionary NAS on a host and a Raspberry Pi endpoint running the online inference (Section~\ref{sec:deployment}).
\item Measurements on the Pi showing that online inference with the single champion and with the ensemble of 40 island champions runs in real time, far inside the daily decision cycle, at low compute and energy cost (Section~\ref{sec:deployment_results}).
\item Evidence that forecasting with a rank-mean ensemble of the island champions earns a much higher net return than forecasting with the single champion used in prior ONE-NAS work (Section~\ref{sec:res-rule}).
\end{itemize}

%% file: 02-RelatedWork.tex
\section{Related Work}

Machine learning for cross-sectional stock return prediction is established \cite{gu2020ml}, and deep models such as MASTER \cite{li2024master} and StockMixer \cite{fan2024stockmixer} set the current benchmarks for daily stock ranking \cite{olorunnimbe2023survey}. They are trained on a fixed span and refitted on a schedule, so they degrade between refits as the market drifts. Incremental learning updates a fixed model each period with meta-learned adaptation of data and parameters \cite{zhao2023doubleadapt}. Online time series forecasting addresses the same concept drift problem \cite{lu2019drift} for general streams: FSNet \cite{pham2022learning} adds fast and slow adaptation paths to a fixed backbone, OneNet \cite{wen2023onenet} learns online combination weights over fixed backbones, and Proceed \cite{zhao2025proactive} adapts parameters before drifted data arrive. In all of these the architecture is fixed before the stream starts and only weights or mixture weights adapt.

Evolutionary NAS searches architectures with a population under selection \cite{stanley2019neuroevolution,liu2021survey}. EXAMM \cite{ororbia2019examm} evolves RNN topologies and cell types from a minimal seed genome with parent-to-offspring weight inheritance, so each candidate needs only brief training; it has been applied to power plant forecasting \cite{lyu2021neuroevolution} and, offline, to stock return prediction, where it matched or outperformed transformer and recurrent baselines with far fewer parameters and no GPU \cite{lyu2025evolving}. ONE-NAS \cite{lyu2022onenas,lyu2023onenas} moves this search online and outperformed online linear, LSTM, GRU and ARIMA baselines on wind turbine and stock index data, reading the population through its single best genome. Other NAS for financial forecasting searches fixed chain-structured spaces offline \cite{levchenko2025chainnas}. Reading an evolved population as an ensemble dates to early neuroevolution \cite{yao1998population}, and neural ensemble search \cite{zaidi2021nes} showed that architecturally diverse ensembles are more robust under dataset shift; neither idea has been combined with online architecture search.

Deployment on constrained hardware is usually handled on the model side. Hardware-aware NAS \cite{wang2021hardware,lin2020mcunet} searches for networks that meet a latency or memory budget, and once-for-all training \cite{cai2019once} extracts a subnetwork per device from one supernet; in both the search finishes before deployment. On-device learning \cite{ren2021tinyol} updates the weights of a fixed network on the device. End-edge-cloud collaborative computing \cite{wang2024end} instead trains on a resourced host and infers on the device that receives the data, shipping the model between them. Our pipeline follows this pattern, except that the host runs a continuous architecture search and ships a new set of small networks each generation rather than new weights for one. Stock prediction models are evaluated by backtest \cite{olorunnimbe2023survey}; serving measurements on edge hardware are rarely reported, and this paper provides them for a Raspberry Pi pilot.

%% file: 03-Method.tex
\section{Method}
\label{sec:method}

\begin{figure}[t]
\centering
\includegraphics[width=\columnwidth]{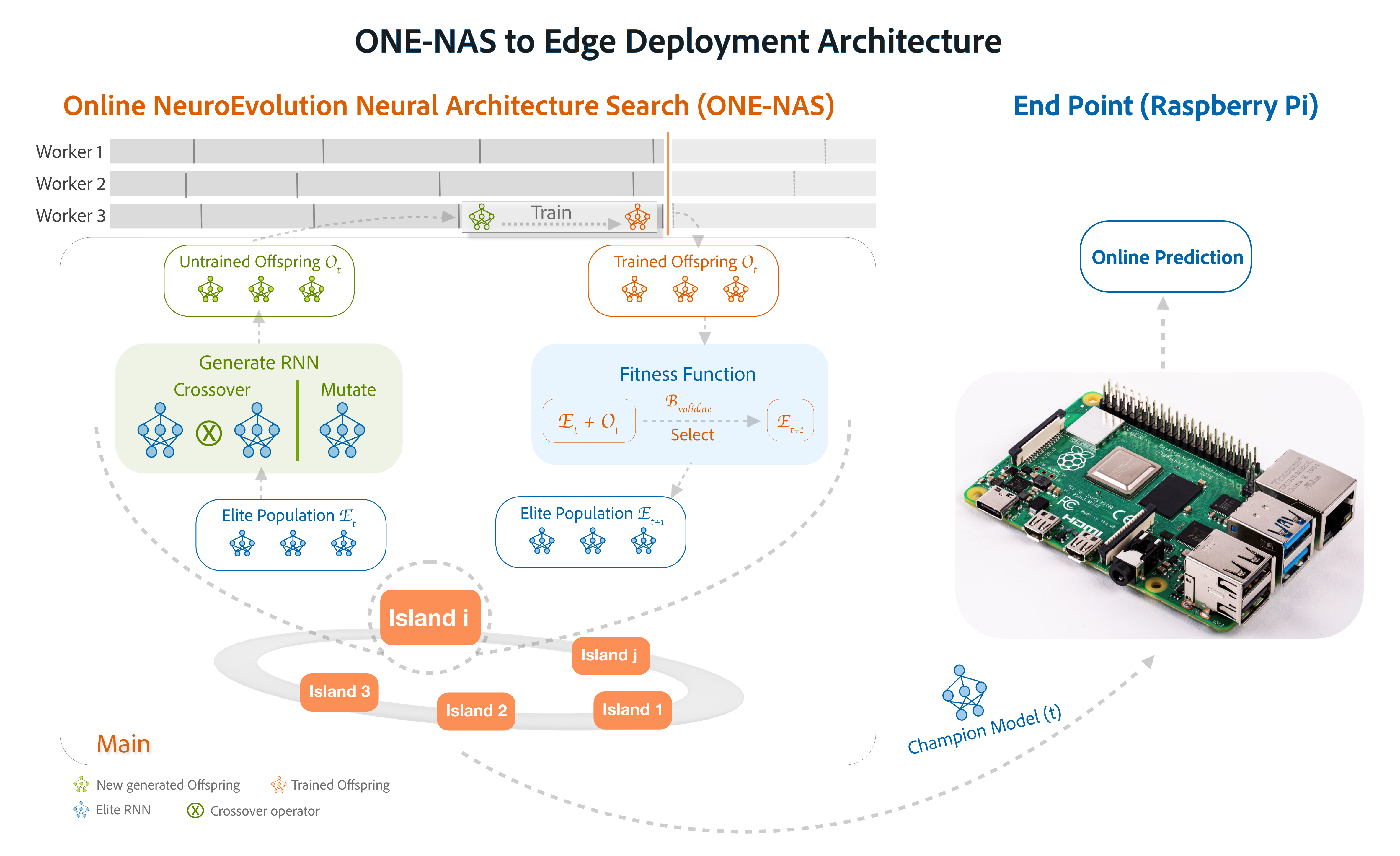}
\caption{Deployment pipeline: ONE-NAS search on the host, online
prediction on the Raspberry Pi endpoint.}
\label{fig:pipeline}
\end{figure}

Online NeuroEvolution-based Neural Architecture Search
(ONE-NAS)~\cite{lyu2023onenas} is a neuroevolution algorithm for
online time series forecasting: it maintains a population of RNNs
whose architectures and weights evolve continuously as new
observations arrive, so the models track a non-stationary stream
without pre-training. This work applies ONE-NAS to stock return
prediction, forecasting the forward return of every security in a
portfolio on each trading day, with the search running on a host
machine and the evolved networks predicting on a separate endpoint.

\subsection{Online Neuroevolution}

ONE-NAS grows networks progressively rather than sampling from a fixed
architecture space. An individual, a genome, is a directed graph
encoding both an architecture and its weights, with hidden nodes drawn
from a library of recurrent cell types and edges that are either
feedforward or recurrent, a recurrent edge carrying a value forward in
time. The search starts from a minimal seed genome holding only
input-to-output connections, so depth, recurrent connectivity and cell
composition are outcomes of the search.

Offspring are produced by mutation (node and edge insertion or
deletion, recurrent edge insertion, and cell type change) and by
crossover between two parents. Weights are inherited for every retained
component and randomly initialised only for new ones, so each offspring
starts from a partially trained network and needs only a few
backpropagation epochs, which keeps the search within the data cadence.
The search runs on $N$ islands, sub-populations that evolve in
parallel, each seeded independently with a minimal genome and growing
as its offspring accumulate. Islands exchange genetic material only
through inter-island crossover at a low rate, so they become
increasingly distinct over time, and a sufficient number of islands is
kept to preserve diversity in the evolutionary search.

A main process holds the islands, generates offspring, and dispatches
each one to the next available worker, which trains and evaluates it
and returns it with its fitness; the main then places it back into
its island and issues the next genome to that worker. Workers are not
synchronised with one another, so a generation's wall-clock cost
depends on the number of workers rather than the population size.

At the end of each generation, every island ranks the genomes it holds
and retains the best performing of them as parents for the next
generation. The best genome on an island is its champion, and the
champion of highest fitness across all islands is the global best
genome. Champions are re-selected every generation, so the networks
that forecast the next window are the current best of the search rather
than a model fixed in advance
(Section~\ref{sec:prediction_rules}).

\subsection{Online Protocol}
\label{sec:online_protocol}

The input stream is partitioned into consecutive windows of $W$ trading
days, with one generation per window (values in
Section~\ref{sec:setup}). At the start of generation $t$, the revealed
history $\mathcal{H}_t$ is split into a \emph{validation block} (its $V$
most recent windows) and a \emph{training pool} (all earlier windows).
Each island generates offspring from its elites and trains them on
subsequences of length $L$ drawn from the training pool alone; all
genomes are then scored by mean squared error on the validation block,
each island keeps its top-ranked genomes as elites, and the genomes
designated by the prediction rule (Section~\ref{sec:prediction_rules})
predict window $t{+}1$ day by day as its features arrive. When window
$t{+}1$ closes, its realised returns are appended to form
$\mathcal{H}_{t+1}$, the validation block slides forward one window, and
generation $t{+}1$ begins. Training draws on the full history while
fitness is measured on the most recent data, so the population tracks
the current regime without discarding earlier ones.

Leakage is prevented by construction: a window is predicted before its
labels are revealed, enters the validation block only afterwards, and
enters the training pool only after leaving it, so no genome is trained
on data it is scored on or selected on data it later predicts. A master
process generates genomes and workers train them asynchronously, so a
generation's wall-clock cost depends on the number of workers rather
than the population size.

\subsection{Prediction Rules}
\label{sec:prediction_rules}

Which genomes predict window $t{+}1$ is a scoring-time decision that
does not affect the search, so the two rules below are compared on the
same runs and any difference between them is attributable to how the
population is used rather than how it was built. Under the
\emph{single champion} rule, used in all prior ONE-NAS work, the
global best genome alone predicts. Under the
\emph{island-champion ensemble} rule, the best elite of each island
predicts, each predicted cross-section is converted to ranks, and the
ranks are averaged across the $N$ members. Ranks rather than raw outputs
are averaged because the members have non-comparable output scales and
every trading rule used here reads only the ordering.

\subsection{Deployment Architecture}
\label{sec:deployment}

Search and inference run on separate machines because population
training is parallel and expensive whereas evaluating the resulting
networks is not (Figure~\ref{fig:pipeline}). The \emph{search host} is
a cloud instance or workstation with enough cores to train a
generation's offspring in parallel; it stores the price history,
advances the window clock, and performs all evolutionary operations.
The \emph{endpoint} is a Raspberry Pi; it performs no training and
stores only the last received genomes and the current day's features.
The Pi is a deliberately constrained target: meeting the data cadence
there bounds the hardware required at the point of use.

After selection in generation $t$, the host serialises the genomes
designated to predict window $t{+}1$ (the global best genome or the
$N$ island champions) into a binary encoding of graph structure, cell
types and weights, and transmits it to the endpoint over TCP/IP. The
endpoint deserialises the networks and predicts each day of window
$t{+}1$ as its features arrive, applying rank aggregation when an
ensemble is received, while the host trains generation $t{+}1$.
Prediction and search thus overlap in time but are decoupled across
hardware.

Inference cost scales only with $N$ and the number of securities, so a
larger island count leaves the endpoint's hardware requirement
unchanged. The endpoint is also stateless with respect to the search:
if the host is unavailable it keeps predicting with the last received
genomes and resumes at the next hand-off.

%% file: 04-Dataset.tex
\section{Dataset and Experimental Setup}
\label{sec:setup}

The ONE-NAS implementation, the trading simulation, and the scripts
that build the panels from CRSP are open source,\footnote{ONE-NAS and
the edge deployment: \url{https://github.com/zimenglyu/ONENAS}; trading
simulation: \url{https://github.com/zimenglyu/Financial_toolbox}; CRSP
panel construction: \url{https://github.com/zimenglyu/CRSP_Processor}.}
and Appendix~\ref{app:details} gives the full experimental detail.

\subsection{Dataset}
\label{subsec:validation-portfolios}

The Center for Research in Security Prices (CRSP) database is the
standard source of U.S. stock prices, returns and volumes in academic
finance, because it is survivorship-bias free and adjusts returns for
splits, dividends and delistings. From it we build four disjoint
portfolios, $V1$ to $V4$, of 50 S\&P~500 stocks each, drawn at random
from the 203 constituents with market capitalisation between 2 and 50
billion USD and at least 20 years of history; each portfolio runs
from early 2004 to the end of 2024.

Each model receives seven daily per-stock inputs, all standard
predictors computed from information available at or
before the prediction day: the daily return and its cross-sectional
rank, bid--ask spread, illiquidity, 20-day prior return, turnover
acceleration and 21-day volatility. The output is the next day's
return, rank-normalised across that day's stocks. A rank is predicted
rather than a raw return because raw returns are noisy and their scale
drifts over two decades, whereas the trading rules act only on which
stocks are expected to do better than which. Feature definitions and
the normalisation are given in Appendix~\ref{app:data}.

\subsection{ONE-NAS Setup}
\label{sec:protocol}

ONE-NAS runs on a weekly clock: each window is $W=5$ trading days,
one generation is executed per window, fitness is MSE on the
validation block, the $V=5$ most recent windows, and offspring train
for 10 backpropagation epochs on subsequences of $L=40$ days. Unless
stated otherwise, every experiment uses $N=40$ islands and all six
recurrent cell types (simple, UGRNN, MGU, GRU, $\Delta$-RNN and
LSTM). The full configuration, how it was selected on the tuning
span, and the fifteen single-axis variants trained are in
Appendix~\ref{app:onenas}.

A ONE-NAS run is one uninterrupted online pass over the data: at every
window the population learns from all history before that window
(Section~\ref{sec:online_protocol}) and is not reset at span
boundaries; the three spans differ only in what the predictions made
along the way are used for. Burn-in (2004--2015) grows the population
from the minimal genome and records no prediction. The tuning span
(2016--2019) is where hyperparameters were selected: candidate
configurations, each an online run of this kind, were compared on
their 2016--2019 predictions, and every configuration decision was
made on this span alone. Runs with the selected configuration then
continue through the evaluation span (2020--2024) with no
configuration change, and results are reported on 2022--2024, the
span of our prior study on these panels \cite{lyu2025evolving}. Each
configuration is run with 10 seeds on each of the 4 panels, and all
comparisons and $t$ statistics are paired within (panel, seed) cells.
Each search ran on one whole 128-core node of the Anvil cluster (NSF
ACCESS)~\cite{song2022anvil}; the inference endpoint is a Raspberry Pi 4B
(Section~\ref{sec:deployment}).

\subsection{Baselines}

ONE-NAS is compared against four baselines that are also online
models: an online LSTM and an online GRU that take one gradient step
per trading day, a periodic LSTM retrained from scratch each month
(the common batch-retraining practice), and an online first-order
autoregressive model. All follow the same prequential rule as ONE-NAS
(predict the next day, reveal the outcome, update) and use the same
inputs, target, trading rules and costs. Each is a single model at the
same scale as the networks ONE-NAS evolves, so the comparison isolates
architecture search and the population rather than model capacity.
Architectures and training settings are in
Appendix~\ref{app:baselines}.

\subsection{Trading Rules and Metrics}

Predictions are turned into positions by three trading rules, each
reading only the ordering of the day's predicted
cross-section: \emph{Overlapping}, which each day buys the 10 top-ranked
stocks, sells short the 10 bottom-ranked, and holds the position for
10 days \cite{efficiency1993returns}; \emph{Banded}, which enters a
stock when it reaches the top 10 and holds it until it falls past rank
35, so that small rank changes do not trigger trades
\cite{novy2016taxonomy}; and \emph{Conditional long--short}, the daily
rebalancing rule of our prior study \cite{lyu2025evolving}.
Transaction costs are charged per stock and per day from the CRSP
bid--ask spread, and all reported returns are net of them. Position
sizing, netting and the cost model are in Appendix~\ref{app:books}.

Prediction quality is measured by the rank IC, the daily Spearman
correlation between predicted and realised returns averaged over
scored days; it measures how well the model orders the stocks each
day, which is what the trading rules act on. Economic performance is
measured by net return after costs, annualised Sharpe ratio and
turnover, and by alpha and market beta from a Fama--French plus
momentum regression \cite{fama1993common,carhart1997persistence}.
Definitions are in Appendix~\ref{app:metrics}.

%% file: 05-Results.tex

\section{Results}
\label{sec:results}

All ONE-NAS hyperparameters were selected on the 2016--2019 tuning
span, disjoint from the evaluation span (Appendix~\ref{sec:res-hp};
full configuration in Table~\ref{tab:hps}). Unless stated otherwise,
ONE-NAS figures are for the island-champion ensemble at $N=40$ islands
trading the Overlapping rule, averaged over the four panels
$V1$--$V4$ and ten seeds ($n=40$ cells) on the 2022--2024 evaluation
span, and paired comparisons run over the (panel, seed) cells the two
arms share. Section~\ref{sec:deployment_results} measures inference
on the endpoint, Sections~\ref{sec:res-headline}
and~\ref{sec:res-rule} show that the method works and where the gain
comes from, Section~\ref{sec:res-robust} checks the trading rule,
transaction costs and factor exposure, and
Section~\ref{sec:res-operating} gives operating guidance for the
pilot. Supporting tables are in Appendix~\ref{app:results}.

\subsection{On-Device Inference Cost}
\label{sec:deployment_results}

\begin{figure}[t]
\centering
\includegraphics[width=0.9\columnwidth]{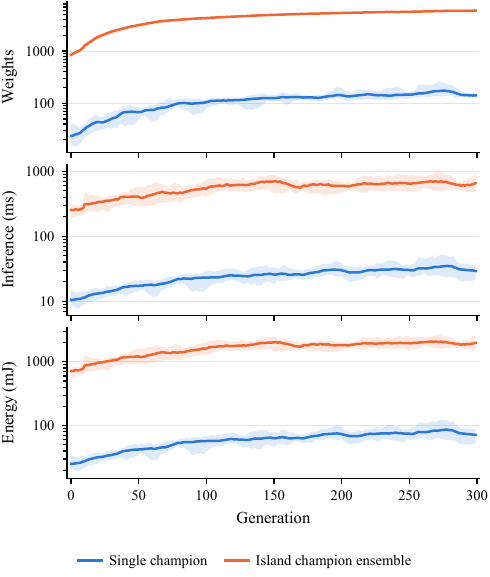}
\caption{On-device cost per pass of the two deployment options, on log
axes; the line is the mean over ten runs and the band the min--max
envelope. The island-champion ensemble runs all 40 island champions,
$39\times$ the weights of the single champion.}
\label{fig:pi-modes}
\end{figure}

The champions evolved online at generation $t$ are sent to the
Raspberry Pi 4B endpoint of Section~\ref{sec:deployment}, which
predicts the test window of generation $t$. Figure~\ref{fig:pi-modes}
shows time and energy per pass for the single champion and for the
island-champion ensemble of all 40 island champions, as the mean and
min--max envelope over ten repeated runs. Energy is whole-board energy
measured with an INA219, with idle power metered at the start of each
run; per-seed and per-quarter values are in Appendix~\ref{app:res-pi}.

Inference is cheap for a single network: a pass over the 50-stock
window takes 24.6~ms, and 88\% of its energy is the board's idle draw,
so at this scale uptime rather than compute dominates. Cost is set by
the architecture rather than the hardware: as the evolved networks
grow over the campaign, time and energy grow with them while board
power stays flat (Table~\ref{tab:pi-progression}). The ensemble
deploys $39\times$ the weights at $23\times$ the time (556~ms per
pass) and is more accurate: its MSE of $0.975$ beats the single
champion's $1.031$ and the average of its own members, so combining
the island champions beats the genome the search would have shipped.
Both options remain far inside a daily decision cycle on a \$50 board;
the choice between them is an accuracy--energy trade rather than a
feasibility one, and the size the search settles on governs both: size
control is also energy control.

\subsection{Net Return against Baselines}
\label{sec:res-headline}

\begin{figure}[t]
\centering
\includegraphics[width=\columnwidth]{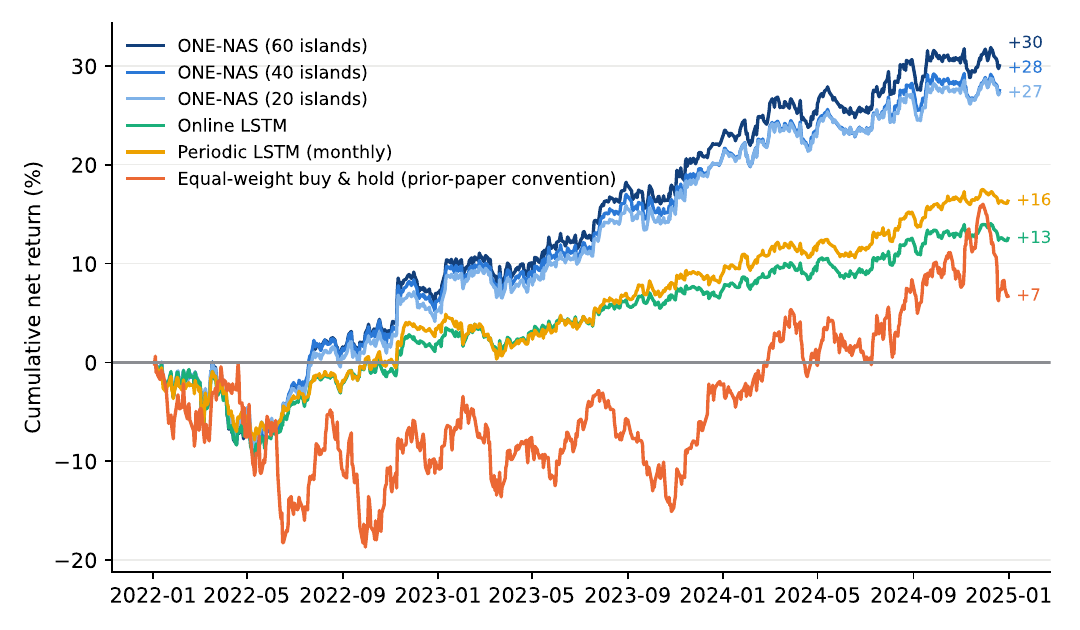}
\caption{Cumulative net return, 2022--2024 (books restarted at window
start). Buy and hold follows the prior-study benchmark convention.}
\label{fig:curves}
\end{figure}

We trade the ensemble at 20, 40 and 60 islands, the four baselines
of Section~\ref{sec:setup} and an equal-weight buy and hold of the
same universe, every arm on the Overlapping rule under the same
realised costs, so the comparison isolates the predictions
(Table~\ref{tab:window}; cumulative paths in
Figure~\ref{fig:curves}).

\begin{table}[t]
\centering
\footnotesize
\begin{tabular}{@{}l@{\hspace{5pt}}r@{\hspace{5pt}}r@{\hspace{5pt}}r@{\hspace{5pt}}r@{\hspace{5pt}}r@{}}
\toprule
 & 2022 & 2023 & 2024 & 2022--24 & $t^{c}$ \\
\midrule
ONE-NAS ($N{=}60$) & $+6.9$  & $+13.7$ & $+7.0$ & $+30.1$\se{0.9} & $-1.63$ \\
ONE-NAS ($N{=}40$) & $+6.1$  & $+12.4$ & $+6.3$ & $+27.5$\se{1.1} & --      \\
ONE-NAS ($N{=}20$) & $+5.1$  & $+13.3$ & $+6.3$ & $+27.4$\se{1.1} & $0.10$  \\
\midrule
Online LSTM        & $+1.2$  & $+4.5$  & $+3.8$ & $+11.3$\se{1.8} & $7.32$  \\
Online GRU         & $+2.1$  & $+3.1$  & $+5.0$ & $+11.7$\se{2.2} & $6.59$  \\
Periodic LSTM (mo.) & $+1.4$ & $+3.9$ & $+6.9$ & $+14.8$\se{1.0} & $6.66$  \\
Online AR$^{a}$    & $-14.6$ & $+4.8$  & $+7.9$ & $-3.3$          & $2.22$  \\
\midrule
Buy \& hold$^{b}$  & $-10.1$ & $+11.2$ & $+6.8$ & $+7.9$          & --      \\
\bottomrule
\end{tabular}
\par\smallskip
{\footnotesize
$^{a}$Deterministic, one run per panel ($n{=}4$). $^{b}$Prior-study
convention; totals the yearly cells, not seed-paired. $^{c}$Paired
$t$ of the 40-island ensemble minus the row over shared (panel, seed)
cells; positive favours the ensemble.\par}

\caption{Net return (\%) by trade year, mean over $V1$--$V4$ and seeds
(10 for ONE-NAS; 48, 40 and 40 for the online LSTM, online GRU and
periodic LSTM, Appendix~\ref{app:baselines}). The 2022--24 column is
one run over the window, not the sum of the yearly cells.}
\label{tab:window}
\end{table}

At 40 islands the ensemble returns $+27.5\%$ against $+11.3$ to
$+14.8\%$ for the three neural baselines, a paired margin of $+12.7$
to $+16.2$ points at $t>6.5$ over 40 cells; 20 and 60 islands are not
separable from it. The margin is not the product
of a single year: the ensemble leads every baseline in 2022 and 2023,
and in 2024 the monthly-retrained LSTM draws level ($+6.9$ against
$+6.3$) while the AR model, which loses money over the window, has its
one good year ($+7.9$). Buy and hold returns $+7.9\%$ but is a market
reference rather than a like-for-like competitor: it runs at a market
beta of $0.80$ and earns no alpha (Table~\ref{tab:alphas}).

\subsection{Prediction Rule: Single Champion versus Island-Champion Ensemble}
\label{sec:res-rule}

Prior ONE-NAS work reads the evolved population through its single
global-best genome; the headline reads it as a rank-mean ensemble of
island champions. Both are scoring-time readings of the same runs, so
Table~\ref{tab:m1} compares them within-run, paired over seeds.
Tables~\ref{tab:diversity} and~\ref{tab:invariance} ask where the
gain comes from: the members' rank IC and pairwise correlation, and
the gain inside all 16 configurations of the tuning screen.

\begin{table}[t]
\centering
\footnotesize
\begin{tabular}{@{}l@{\hspace{4pt}}r@{\hspace{4pt}}r@{\hspace{4pt}}r@{\hspace{4pt}}r@{}}
\toprule
Fleet & Single & Ensemble & $\Delta$net ($t$) & $\Delta$Sharpe ($t$) \\
\midrule
60 isl.\           & $+1.0$ / $0.09$ & $+30.1$ / $0.93$ & $+29.1$ $(18.3)$ & $+0.84$ $(14.1)$ \\
40 isl.\           & $+4.5$ / $0.21$ & $+27.5$ / $0.87$ & $+23.0$ $(18.2)$ & $+0.66$ $(14.7)$ \\
20 isl.\           & $+4.7$ / $0.25$ & $+27.4$ / $0.87$ & $+22.6$ $(14.9)$ & $+0.62$ $(14.9)$ \\
\midrule
40 isl.$^{a}$      & $+8.9$ / $0.30$ & $+36.7$ / $1.22$ & $+27.8$ $(7.5)$  & $+0.92$ $(6.6)$  \\
\bottomrule
\end{tabular}
\par\smallskip
{\footnotesize $^{a}$Replication on the 2016--2019 tuning-span fleets,
at $n{=}5$ seeds; every other row is the 2022--2024 trading span at
$n{=}10$.\par}
\caption{Prediction rule, within-run paired: single global-best genome
vs.\ island-champion rank-mean ensemble from the same runs (Overlapping
rule). Net \% / Sharpe; $\Delta$ = ensemble $-$ single, paired $t$ over
seeds.}
\label{tab:m1}
\end{table}

The single champion returns $+4.5\%$ at 40 islands at Sharpe $0.21$,
below every neural baseline in Table~\ref{tab:window}; the same runs
read as an ensemble return $+27.5\%$ at Sharpe $0.87$, a paired gain
of $+23.0$ points ($t=18.2$), and the tuning-span fleets replicate it
($+8.9$ against $+36.7$, $t=7.5$). The search does not pay by finding
one good architecture; it pays by producing a population. The size of
the gain is what diversity implies: individual champions are weak
(mean rank IC $+0.0065$ to $+0.0068$) and weakly correlated (mean
pairwise Spearman $\rho$ of $0.150$ to $0.166$), and the
equicorrelated averaging prediction $m\sqrt{N/(1+(N-1)\rho)}$ matches
the measured ensemble IC within 3\%. The gain is also not specific to
the tuned configuration: it is positive in 16 of 16 configurations of
the tuning screen (mean $+8.3$ points; smaller because those cells run
at the 8-island registered width).

\subsection{Trading Rule, Transaction Costs and Factor Exposure}
\label{sec:res-robust}

Three checks ask whether the headline depends on how the predictions
are traded or on the cost assumption, and whether it is factor
exposure in disguise: the Banded and Conditional long--short rules of
Section~\ref{sec:setup} on the same 40-island predictions and costs
(Table~\ref{tab:books}); the realised per-name cost scaled by
$2\times$, $3\times$ and $5\times$ with its cross-sectional shape
preserved (Table~\ref{tab:costs}); and each arm's pooled daily book
return regressed on the Fama--French three factors plus momentum
(Table~\ref{tab:alphas}).

\begin{table}[t]
\centering
\footnotesize
\begin{tabular}{@{}l@{\hspace{3pt}}r@{\hspace{3pt}}r@{\hspace{3pt}}r@{\hspace{3pt}}r@{\hspace{3pt}}r@{}}
\toprule
Arm & $1\times$ & $2\times$ & $3\times$ & $5\times$ & B/E$^{a}$ \\
\midrule
ONE-NAS ($N{=}60$)        & $+30.1$ & $+27.2$ & $+24.4$ & $+18.8$ & $11.6\times$ \\
ONE-NAS ($N{=}40$)        & $+27.5$ & $+24.7$ & $+21.8$ & $+16.0$ & $10.6\times$ \\
ONE-NAS ($N{=}20$)        & $+27.4$ & $+24.4$ & $+21.5$ & $+15.6$ & $10.3\times$ \\
Periodic LSTM              & $+14.8$ & $+11.4$ & $+8.0$  & $+1.2$  & $5.4\times$ \\
Online GRU                 & $+11.7$ & $+8.3$  & $+4.9$  & $-1.8$  & $4.5\times$ \\
Online LSTM                & $+11.3$ & $+8.1$  & $+4.9$  & $-1.5$  & $4.5\times$ \\
Online AR                  & $-3.3$  & $-6.0$  & $-8.8$  & $-14.2$ & -- \\
\bottomrule
\end{tabular}
\par\smallskip
{\footnotesize
Scaling multiplies the realised per-name cost, so the cross-sectional
shape of costs is preserved. The $1\times$ column is the 2022--24 column
of Table~\ref{tab:window}. $^{a}$Break-even: the cost multiple at which
net return reaches zero.\par}
\caption{Cost sensitivity: net return (\%) on 2022--2024 as the realised
per-name transaction cost is scaled by the stated multiple.}
\label{tab:costs}
\end{table}

Banded returns $+33.9\%$ and Conditional long--short $+25.0\%$
against $+27.5\%$ for Overlapping, but Banded is uneven across years
($+12.5$, $+14.6$, $+3.4$) with larger seed dispersion (SE $1.8$
against $1.1$) and Conditional long--short reduces to a plain daily
rebalance on rank-normal predictions, so Overlapping is the rule used
throughout. The ensemble stays profitable to $10$--$12\times$
realistic costs against $4.5$--$5.4\times$ for the neural baselines,
and at $5\times$ still returns $+15.6$ to $+18.8\%$ while the online
LSTM and GRU are below zero; part of the margin is lower turnover,
$0.12$ of gross book per day against $0.13$--$0.14$. After factor
adjustment the three ensemble widths carry alphas of $+10.0$ to
$+11.2\%$ per year ($t=2.56$ to $2.77$, Newey--West lag 10) at market
betas of $0.11$--$0.12$, and only the monthly-retrained LSTM among the
baselines reaches $|t|=2$: the book is close to market-neutral and its
return is not factor exposure.

\subsection{Operating the Pilot}
\label{sec:res-operating}

\begin{figure}[t]
\centering
\includegraphics[width=\columnwidth]{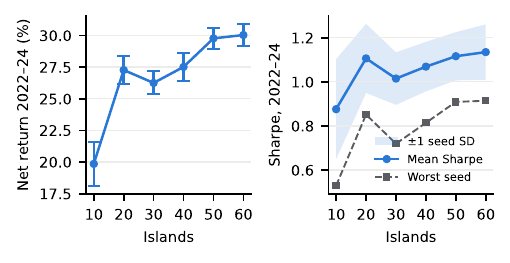}
\caption{Performance vs.\ island count on the evaluation span
(2022--2024, Overlapping rule; 10 seeds $\times$ 4 panels at every
width). Left: pooled net return (mean $\pm$ SE). Right: Sharpe, mean
with $\pm 1$ seed-SD band, and the worst seed. The 40-island headline
was fixed on the tuning span (Table~\ref{tab:tunewidth},
Appendix~\ref{sec:res-hp}) before this curve was measured.}
\label{fig:islands}
\end{figure}

A deployment holds a single run and must decide when to start its
population and how many islands to run. We compare a population
evolving online since 2020-01 with one cold-started at 2022-01, both
scored on 2022--2024 at the frozen 8-island registered configuration,
and run island counts from 10 to 60 at ten seeds and four panels on
the evaluation span (Figure~\ref{fig:islands}).

The continuously evolving population returns $+26.5\%$ against
$+12.3\%$ for the cold start: more than half of the return at this
configuration comes from the two years of evolution before the
evaluation span, so a deployed population should be started well ahead
of the period it trades and never reset. Return and Sharpe rise
steeply from 10 to about 20 islands and then flatten, and 20 to 60
islands are not separable at ten seeds; over the same range the
standard deviation of per-seed Sharpe falls from $0.227$ to
$0.11$--$0.13$ and the worst seed rises from $0.53$ to $0.91$. Width
buys consistency at least as much as expected return, at an inference
cost on the endpoint that scales with $N$
(Section~\ref{sec:deployment_results}). 50 and 60 islands score
nominally above 40 on the trading period but are indistinguishable
from it on the tuning span (Appendix~\ref{sec:res-hp}) and are not
adopted.

%% file: 06-Conclusion.tex
\section{Conclusion}
\label{sec:conclusion}
This work proposed a host and endpoint pipeline that deploys online
neuroevolutionary NAS for stock forecasting and for trading on the
forecasts: ONE-NAS evolves the population on a search host and ships
each generation's champions to a Raspberry Pi, which forecasts online.
Results on the Pi show that the deployment can be used in practice:
the single champion predicts a 50-stock window in 24.6~ms and the
ensemble of 40 island champions in 556~ms, far inside the daily
cadence and at low energy.

This work also explored two ways of using the evolved population for
prediction, the single champion of prior ONE-NAS work and the
island-champion ensemble. Results show that the ensemble is far
better: on four panels of US mid-cap equities over 2022--2024 it
returns $+27.5\%$ net of costs against $+4.5\%$ for the single
champion, a paired gain of $+23.0$ points ($t=18.2$) at no search-time
cost, positive in every configuration we trained, and it outperforms
online LSTM, online GRU and monthly-retrained LSTM baselines, which
return $+11.3$ to $+14.8\%$ under the same rule and costs ($t>6.5$).
Future work includes the full deployment: adding an execution layer
and live measurement of realised costs, and having users in the
industry run the pipeline and give feedback.

%% file: 80-Appendix.tex
\providecommand{\text}[1]{\mathrm{#1}}

\appendix
\section{Experimental Details}
\label{app:details}

This appendix gives the detail summarised in Section~\ref{sec:setup}
and follows its order. Appendix~\ref{app:data} describes the panels,
features and target of Section~\ref{subsec:validation-portfolios};
Appendix~\ref{app:onenas} lists the ONE-NAS configuration of
Section~\ref{sec:protocol}, the tuning-span screen behind it, and the
pre-registration under which it was scored; Appendix~\ref{app:baselines}
gives the baseline architectures; Appendix~\ref{app:books} gives the
position sizing, netting and cost model of the three trading rules,
with the Conditional long--short rule as pseudocode; and
Appendix~\ref{app:metrics} defines the metrics and the paired tests.
Appendix~\ref{app:results} then reports the hyperparameter selection
on the tuning span and the tables that support
Section~\ref{sec:results}. The main text carries every number it
relies on; this appendix documents how those numbers were produced.

\subsection{Dataset Construction}
\label{app:data}

The universe is the set of S\&P~500 constituents in the CRSP daily
panel with current market capitalisation $2 \le M_i < 50$ billion USD
and at least 20 years of continuous trading history. Mid-cap stocks
receive less analyst and institutional scrutiny than large-caps and
are therefore less efficiently priced. The restriction leaves 203
names, from which 200 were drawn at random, shuffled, and split evenly
into $V1$--$V4$, so no stock appears in more than one portfolio. Each
portfolio is stored as one series per ticker with a panel-level date
index carrying the aligned price and transaction-cost columns. All
four end on 2024-12-31 and each begins once all 50 constituents have
priced history: $V1$ from 2004-08-16, $V2$ from 2004-12-17, $V3$ from
2004-10-27 and $V4$ from 2004-02-09, giving 5{,}130, 5{,}043, 5{,}079
and 5{,}257 trading days. The differing lengths affect only burn-in;
the tuning and evaluation spans are the same calendar spans for all
four, so no method gains or loses scored days.

The seven features are computed causally, from information available
at or before the day on which a position is formed. Three describe
the stock's recent price behaviour, three describe how easily and how
heavily it trades, and one places it against the rest of the market
that day. Liquidity and attention features are included because the
target is short-horizon: over a single day, how expensive a stock is
to trade and how much attention it is attracting carry more
information about its next move than fundamentals do, and they also
tell the model where a predicted edge would be eaten by trading
costs. In the order of Section~\ref{subsec:validation-portfolios},
the features are:

\begin{itemize}\itemsep2pt
\item \texttt{RET}: the raw daily return.
\item \texttt{RET\_CS\_IN}: the cross-sectional rank-normal score of
      \texttt{RET} on the same day,
      $\Phi^{-1}\!\big((\mathrm{rank}_t(i)-0.5)/N_t\big)$ over the
      $N_t$ names priced that day.
\item \texttt{BA\_SPREAD}: the quoted bid--ask spread, the gap between
      the best buying and selling price, a proxy for how costly the
      stock is to trade.
\item \texttt{ILLIQUIDITY}: Amihud illiquidity, absolute return over
      dollar volume, smoothed over 21 days and logged. It is large when
      a small amount of trading moves the price a lot.
\item \texttt{REV21\_1}: the 20-day prior return, that is, the
      compounded return over the 20 days before the formation day,
      excluding it. Recent winners tend to give some of it back, so
      this is informative with a negative sign.
\item \texttt{TURN\_RATIO}: turnover acceleration, the mean of
      $\log(1+\mathrm{turnover})$ over the last 5 days minus the same
      mean over the last 63. It rises when a stock begins attracting
      unusual attention.
\item \texttt{VOL21}: realised volatility, the standard deviation of
      \texttt{RET} over the trailing 21 days.
\end{itemize}

The last five are additionally mapped, per date, to their
cross-sectional rank on $[-1,1]$. Without this a feature would carry a
level that drifts across two decades (spreads and volumes in 2004 are
not comparable to 2024) and the model would spend capacity tracking
that drift instead of the cross-section. The prediction target
\texttt{RET\_CS} is the next day's return, rank-normalised across the
cross-section of that day by the same $\Phi^{-1}$ transform, which
makes the target stationary by construction and matches how the
predictions are consumed: every trading rule of
Appendix~\ref{app:books} reads only the ordering of the predicted
cross-section. Realised returns used for scoring are CRSP
\texttt{RET}, adjusted for splits and dividends; prices and per-name
transaction costs enter only through the cost model of
Appendix~\ref{app:books}.

\subsection{ONE-NAS Configuration and Hyperparameter Search}
\label{app:onenas}

\begin{table}[t]
\centering
\footnotesize
\begin{tabular}{@{}l@{\hspace{4pt}}l@{}}
\toprule
Setting & Value \\
\midrule
Islands $N$ & 40 (headline); 8 (reg.) \\
Elites / island & 8 \\
Offspring / island / generation & 5 \\
Mut.\ / intra- / inter-island crossover & 0.7 / 0.2 / 0.1 \\
Repopulation & every 50 generations \\
Fitness & validation MSE \\
Prediction rule & island-champion ensemble \\
Target & next-day CS rank-normal \\
Seq.\ len.\ $L$ / window $W$ / val.\ $V$ & 40 / 5 / 5 \\
Train.\ subseq.\ / genome (NTS) & 2000 \\
BP epochs / genome & 10 \\
Replay (PER) $\alpha$, $\lambda$, $\epsilon$ & 0.6, 0.007, $10^{-8}$ \\
Generations / window (rounds) & 1 \\
Cell types & simple, UGRNN, MGU, \\
 & GRU, $\Delta$-RNN, LSTM \\
Warm-up generations & 50 \\
\bottomrule
\end{tabular}
\caption{ONE-NAS configuration used in Section~\ref{sec:results}, in
the notation of Sections~\ref{sec:online_protocol}
and~\ref{sec:protocol}, selected on the 2016--2019 tuning span
(Appendix~\ref{sec:res-hp}). ``reg.'': the registered 8-island
configuration; abbreviations in parentheses as in
Tables~\ref{tab:hpscreen} and~\ref{tab:invariance}.}
\label{tab:hps}
\end{table}

Table~\ref{tab:hps} lists every setting of the configuration used in
Section~\ref{sec:results}; the prediction rule is the island-champion
rank-mean ensemble of Section~\ref{sec:prediction_rules} and the
target is \texttt{RET\_CS} of Appendix~\ref{app:data}. Each of the
$N$ islands keeps an elite set
of 8 genomes and generates 5 offspring per generation, each offspring
by mutation with probability 0.7, by crossover between two elites of
the same island with probability 0.2, and by crossover with an elite
of another island with probability 0.1. Every 50 generations the
island whose best elite has the worst fitness is cleared and
repopulated from the global best genome, which counters stagnation.
Offspring are trained for 10 backpropagation epochs on 2000
subsequences of length $L=40$ days drawn from the training pool by
prioritised experience replay (PER) rather than uniformly: each
window in the pool carries an error estimate updated from the elite
population's validation MSE, its priority is
$(\mathrm{MSE}+\epsilon)^{-1}e^{-\lambda a}$ with $a$ its age in
generations, and windows are drawn without replacement with
probability proportional to priority$^{\alpha}$, so recent windows and
windows the current elites fit well are drawn more often.

Fitness is MSE on the $V=5$ validation windows, one generation (one
round of offspring generation, training and selection) runs per
$W=5$-day window, and predictions are recorded only after 50 warm-up
generations.

\paragraph{Pre-registration.} The configuration to be scored, the
evaluation span, the trading rule and the metrics were committed to
the repository before the evaluation span was scored; every later
analysis of evaluation-span data was declared in advance with its
decision rule, and any configuration change required a dated
amendment written before the changed configuration was scored. The
configuration committed first, with $N=8$ islands, is the
\emph{registered} configuration referred to in
Section~\ref{sec:res-operating} and Table~\ref{tab:invariance}; the
move to 40 islands was recorded as such an amendment on the
tuning-span evidence of Appendix~\ref{sec:res-hp}, before the
evaluation span was scored. The registered protocol, the full screen
with per-cell outcomes, and all declared exploratory analyses are in
the code release cited in Section~\ref{sec:setup}.

\paragraph{Spans and warm-up.} A run is one online pass from 2004 to
the last scored day (Section~\ref{sec:protocol}). Burn-in
(2004--2015) records no prediction; the tuning span (2016--2019) is
scored to select the configuration; runs with the selected
configuration continue through the evaluation span (2020--2024) with
no configuration change, so evolution is uninterrupted from 2004 to
the last scored day, and 2020--2021 is scored but not reported so that
the reported span matches the prior study on these panels. In both
scored spans 50 warm-up generations precede the first recorded
prediction, so the scored population is never a cold start.

\paragraph{Compute.} All searches ran on the Anvil cluster (NSF
ACCESS) named in Section~\ref{sec:protocol}. Each run occupied one
node with two 64-core AMD EPYC 7763 processors (128 cores,
2.45\,GHz) and 256\,GB of memory, using 128 MPI ranks: one master
maintaining the island populations and generating offspring, and 127
workers training and evaluating them. The island count is independent
of the rank count, so the worker pool is identical across all reported
configurations. Baselines are single-process and ran on the same
hardware. The inference endpoint is the Raspberry Pi 4B (8\,GB RAM) of
Section~\ref{sec:deployment}, measured in
Section~\ref{sec:deployment_results}.

\subsection{Baselines}
\label{app:baselines}

All four baselines are evaluated on the same panels with the same
prequential protocol, features, target, trading rules and transaction
costs as ONE-NAS (Section~\ref{sec:setup}). Each is a single model,
and reported figures are means over independent runs rather than the
return of an ensemble of those runs.

\begin{itemize}\itemsep2pt
\item \textbf{Online LSTM.} A single-layer LSTM with 8 hidden units
      and input sequence length 10, trained with Adam (learning rate
      $10^{-3}$), taking one gradient step per trading day on a
      minibatch drawn from a replay buffer of the most recent 1000
      days. 48 seeds per panel.
\item \textbf{Online GRU.} Identical to the online LSTM with a GRU cell
      and sequence length 5. 40 seeds per panel.
\item \textbf{Periodic LSTM.} The same LSTM architecture, retrained
      from scratch at the start of each month for up to 4000 steps and
      held fixed between retrainings. 40 seeds per panel.
\item \textbf{Online AR.} A first-order autoregressive model fitted by
      online gradient descent with learning rate $10^{-4}$.
      Deterministic, so one run per panel.
\end{itemize}

\subsection{Trading Rules}
\label{app:books}

Every rule reads only the ordering of each day's predicted
cross-section, so all reported metrics are invariant to any strictly
monotone transform of a model's output. Overlapping is the rule used
throughout Section~\ref{sec:results}; Banded and Conditional
long--short are run on the same predictions in
Section~\ref{sec:res-robust} (Table~\ref{tab:books}).

\paragraph{Overlapping.} Following the overlapping-portfolio
construction of \citet{efficiency1993returns}, each day a new
portfolio is formed that goes long the $k=10$ highest-ranked names and
short the 10 lowest-ranked, with equal notional per name, and is held
for $H=10$ trading days; the portfolio formed $H$ days earlier is
closed. Ten portfolios are therefore open at any time, each funded
with $1/H$ of the capital. Positions are netted across portfolios
before trading, so a name that appears in several consecutive
portfolios is held rather than repeatedly bought and sold, and
transaction costs are charged only on the change in aggregate notional
per name.

\paragraph{Banded.} Following the buy/hold spread of
\citet{novy2016taxonomy}, a name is entered when its predicted rank
reaches the top 10 and is held until its rank falls past 35. Names
between the two thresholds are neither bought nor sold, which removes
the churn caused by small day-to-day changes in rank near the
boundary.

\begin{algorithm}[!t]
\caption{Conditional long--short rule, the daily long--short strategy
of \citet{lyu2025evolving}; $K_L = K_S = k = 10$ in this paper}
\label{alg:long-short}
\begin{algorithmic}[1]
\footnotesize
\Function{LongShortReturn}{$P$, $C_0$, $K_L$, $K_S$}
    \State $C \gets C_0$
    \For{$t$ in $[1,T]$}
        \State $S \gets \textsc{SortByPrediction}(P, t)$
        \Comment{\emph{descending}}
        \If{$S[K_L].\hat{y}_t > 0$ \textbf{and} $S[-K_S].\hat{y}_t < 0$}
            \State $C \gets C + \textsc{ClearHoldings}(P, t)$
            \Comment{\emph{close all}}
            \State $q_L \gets C/K_L$, \enspace $q_S \gets C/K_S$
            \For{$i$ in $[1,K_L]$}
                \State $S[i].\textsc{Buy}(q_L, t)$, \enspace $C \gets C - q_L$
            \EndFor
            \For{$i$ in $[1,K_S]$}
                \State $S[-i].\textsc{Short}(q_S, t)$, \enspace $C \gets C + q_S$
            \EndFor
        \EndIf
    \EndFor
    \State $C \gets C + \textsc{ClearHoldings}(P, T)$
    \State \Return $(C - C_0)/C_0$
    \Comment{\emph{net return}}
\EndFunction
\end{algorithmic}
\end{algorithm}

\paragraph{Conditional long--short.} The daily long--short rule of
the prior study on these panels \cite{lyu2025evolving}, given in full
as Algorithm~\ref{alg:long-short}. $P$ is the panel (the predictions
$\hat{y}_t$, prices and per-name costs of every name on every day),
$C_0$ the initial capital, $C$ the cash held, and $T$ the last day of
the test period. Each day $t$ the names are sorted by predicted return
in descending order, $S[i]$ denoting the $i$-th name from the top and
$S[-i]$ the $i$-th from the bottom. If the $K_L$-th highest prediction
is positive and the $K_S$-th lowest is negative, all existing
positions are closed and the capital is redeployed equally, $C/K_L$
long in each of the top $K_L$ names and $C/K_S$ short in each of the
bottom $K_S$, so that the short notional equals the capital;
otherwise the current positions are held. \textsc{Buy},
\textsc{Short} and \textsc{ClearHoldings} trade at the previous close
and charge the per-name cost below on the notional traded, and the
final \textsc{ClearHoldings} marks the book to cash, so the value
returned is the net return of Appendix~\ref{app:metrics}. The
condition is a confidence gate that suspends rebalancing when the
model does not separate the top and bottom of the cross-section by
sign. On rank-normalised predictions the gate is satisfied every day,
so the rule reduces to daily full rebalancing, which is why it adds
nothing over a plain daily rebalance in Section~\ref{sec:res-robust}.

\paragraph{Capital and costs.} Each rule is funded with \$100 of
capital ($C_0$ in Algorithm~\ref{alg:long-short}) and holds \$200 of
gross notional when fully invested. Each stock has a per-share cost
$c_{i,t}$ in dollars, the \texttt{TRAN\_COST} column of the panel
(Appendix~\ref{app:data}), which is the half bid--ask spread from
CRSP: the cost of crossing the spread once, in one direction. Dividing
by the share price converts it to a cost per dollar traded, applied to
the notional actually traded at the previous close, so a trade of
$\$v$ in stock~$i$ pays
\begin{equation*}
  \text{cost} \;=\; v \cdot \frac{c_{i,t}}{|P_{i,t}|},
\end{equation*}
a few basis points for the median name in these portfolios and
considerably more for the least liquid ones. Costs are per-stock and
time-varying rather than a single flat basis-point assumption, so
illiquid names are penalised as they should be. All reported returns
are net of these costs; the cost multiples of Table~\ref{tab:costs}
scale $c_{i,t}$ for every name and day by the stated factor.

\subsection{Metrics and Statistical Comparison}
\label{app:metrics}

The rank information coefficient (rank IC) is the Spearman
correlation between predicted and realised returns across the names
on a given day, averaged over all scored days. Net return is the
cumulative daily profit and loss after transaction costs as a fraction
of initial capital, $(C - C_0)/C_0$ in Algorithm~\ref{alg:long-short}.
The annualised Sharpe ratio is the mean daily return divided by its
standard deviation and scaled by $\sqrt{252}$, a measure of return per
unit of risk. Turnover is the mean daily traded notional as a fraction
of gross book size, which determines exposure to transaction costs.
Maximum drawdown, reported for the tuning span in
Table~\ref{tab:tunewidth}, is the largest peak-to-trough decline in
cumulative net return over the period.

To test whether returns are explained by exposure to known sources of
market risk, each arm's pooled daily book return is regressed on the
Fama--French three factors \cite{fama1993common} and the momentum
factor \cite{carhart1997persistence}. The intercept (alpha) is the
return not attributable to those factors, and the market coefficient
(beta) is the sensitivity to overall market movement. Standard errors
are Newey--West \cite{newey1986simple} with lag 10 to account for the
autocorrelation induced by the 10-day holding period
(Table~\ref{tab:alphas}).

The unit of replication is one (panel, seed) cell. Unless stated
otherwise, each ONE-NAS configuration is run with 10 seeds on 4
panels, giving 40 cells; baselines were run at the higher replication
stated in Appendix~\ref{app:baselines} and their means are reported at
that replication. Comparisons between arms are paired within (panel,
seed) cells so that panel and seed effects cancel, and reported $t$
statistics are paired $t$-tests over those cells. When two arms do not
share a seed set, the pairing is restricted to their common cells and
$n$ is stated.

%% file: 81-AppendixResults.tex
\section{Additional Results}
\label{app:results}

This appendix supports Section~\ref{sec:results}.
Appendix~\ref{sec:res-hp} reports the tuning-span screen on which the
configuration of Table~\ref{tab:hps} was fixed, and
Appendices~\ref{app:res-pi} to~\ref{app:res-robust} hold the tables
behind Sections~\ref{sec:deployment_results}, \ref{sec:res-rule}
and~\ref{sec:res-robust}, in that order.

\subsection{Hyperparameter Selection on the Tuning Span}
\label{sec:res-hp}

\begin{table}[t]
\centering
\footnotesize
\begin{tabular}{@{}r@{\hspace{5pt}}r@{\hspace{5pt}}r@{\hspace{5pt}}r@{\hspace{5pt}}r@{\hspace{5pt}}r@{}}
\toprule
Islands & $n$ & rank IC & Net\% & Sharpe & MDD\% \\
\midrule
8  & 20 & $+0.0132$ & $+26.0$ & $0.92$ & $11.6$ \\
16 & 20 & $+0.0149$ & $+31.5$ & $1.11$ & $10.2$ \\
20 & 20 & $+0.0157$ & $+36.0$ & $1.24$ & $10.4$ \\
40 & 40 & $+0.0169$ & $+36.2$ & $1.21$ & $10.4$ \\
50 & 40 & $+0.0170$ & $+36.4$ & $1.21$ & $10.4$ \\
60 & 40 & $+0.0168$ & $+35.7$ & $1.19$ & $10.8$ \\
\bottomrule
\end{tabular}
\par\smallskip
{\footnotesize
$n$ is (panel, seed) cells; widths 8--20 were run at seeds 42--46 and
40--60 at 42--51. Restricting every width to the shared seeds 42--46
leaves the ordering unchanged.\par}
\caption{Island count on the tuning span, 2016--2019, Overlapping
rule, island-champion ensemble: the span on which the headline width
was chosen. The evaluation-span curve is Figure~\ref{fig:islands}.}
\label{tab:tunewidth}
\end{table}

Every ONE-NAS setting used in Section~\ref{sec:results} was selected
on the 2016--2019 tuning span, disjoint from the 2020--2024 evaluation
span on which results are reported for 2022--2024. The full
configuration is Table~\ref{tab:hps}; the fifteen single-axis variants
trained around it, and the rule used to adopt or reject a change, are
described in Appendix~\ref{app:onenas}, and their per-cell outcomes
are in the code release. The one setting swept in detail is the
island count $N$, because it sets two things at once: the parallelism
of the search, and the size of the island-champion ensemble that
predicts, since every island contributes one champion
(Section~\ref{sec:prediction_rules}). The sweep asks whether the
tuning objective keeps improving with width and where the headline
configuration should be set. Six widths were trained, 8, 16, 20, 40,
50 and 60 islands, each on all four panels; widths 8 to 20 were run
at seeds 42--46 (20 cells) and widths 40 to 60 at seeds 42--51 (40
cells). Every cell trades the Overlapping rule under realised costs
and is read through the island-champion ensemble, as in
Section~\ref{sec:results}.

Table~\ref{tab:tunewidth} reports the sweep. Rank IC rises from
$+0.0132$ at 8 islands to $+0.0149$ at 16 and $+0.0157$ at 20, and is
flat above that: the 40-, 50- and 60-island cells span $+0.0168$ to
$+0.0170$ and are not separable from one another (paired against 40
islands, the largest $|t|$ is $0.23$). Net return and Sharpe follow
the same shape, rising from $+26.0\%$ and $0.92$ at 8 islands to
$+36.0\%$ and $1.24$ at 20, and then holding between $+35.7$ and
$+36.4\%$ and between $1.19$ and $1.24$ from 20 to 60 islands.
Maximum drawdown stays between $10.2$ and $11.6\%$ at every width and
is largest at 8 islands, so the return gained from width is not bought
with deeper drawdowns. Since the narrow and wide fleets were run at
different seed counts, the comparison was repeated on the five seeds
shared by every width, and the ordering is unchanged.

The 40-island headline therefore sits on the plateau of the tuning
objective, and the 8-island setting of the original registration
(Appendix~\ref{app:onenas}) is measurably below it, by $0.0037$ in
rank IC and $10.2$ points of net return. The 50- and 60-island cells
are not adopted: on the only span where a configuration choice can
legitimately be made they are indistinguishable from 40 islands, and
each additional island adds one network to the ensemble the endpoint
has to evaluate (Section~\ref{sec:deployment}). The choice was made
before the evaluation span was scored. The evaluation-span curve of
Figure~\ref{fig:islands}, on which 50 and 60 islands score nominally
above 40, was measured afterwards and did not inform it
(Section~\ref{sec:res-operating}).

The other axes of the screen left the configuration unchanged.
Replacing prioritised experience replay by uniform sampling shows no
advantage on the tuning objective ($\Delta$IC $-0.0024$, $t=-1.9$), so
prioritised replay is kept, and no other single-axis variant met the
paired-improvement bar of Appendix~\ref{app:onenas}, so the control
value of every remaining setting is the one reported. The ensembling
gain inside each of these sixteen configurations is given in
Table~\ref{tab:invariance} (Appendix~\ref{app:res-rule}).

\subsection{On-Device Measurements}
\label{app:res-pi}

Tables~\ref{tab:pi-per-seed} to~\ref{tab:pi-deployment} give the
measurements behind Figure~\ref{fig:pi-modes}
(Section~\ref{sec:deployment_results}), taken on the Raspberry Pi 4B
endpoint of Section~\ref{sec:deployment}: per-seed and per-quarter
values for the single champion (Tables~\ref{tab:pi-per-seed}
and~\ref{tab:pi-progression}), per-seed values for the island-champion
ensemble (Table~\ref{tab:pi-island}), and the two deployment options
side by side (Table~\ref{tab:pi-deployment}). A pass is one prediction
of the 50-stock window; energy is whole-board energy over the metered
window, and the bracketed figures are the marginal cost above the
board's idle draw, which is metered at the start of each run. The mean
rows are the figures quoted in the main text: 24.6~ms and MSE $1.031$
for the single champion, and MSE $0.975$ for the ensemble.

The ensemble runs every island's champion and averages their
predictions, so its weight count and cost are totals over the 40
networks: $39\times$ the weights of the single champion at $23\times$
the time and $27\times$ the whole-board energy per pass, for an MSE of
$0.975$ against $1.031$, lower than every member's own
(Table~\ref{tab:pi-island}, last column). The growth row of
Table~\ref{tab:pi-progression} is the basis for the statement that
inference cost is set by the evolved architecture rather than by the
board. The marginal cost above idle is reported for completeness; the
two campaigns ran on different power supplies, so the marginal ratio
is not quoted.

\begin{table}[t]
\centering
\footnotesize
\setlength{\tabcolsep}{2pt}
\begin{tabular}{@{}lrrrrrrr@{}}
\toprule
Seed & Gen. & Weights & Infer. & Energy & Power & Idle & MSE \\
 & & (end) & (ms) & (mJ) & (mW) & (mW) & \\
\midrule
42 & 300 & 143 & 22.8 & 56.9 & 2507 & 2103 & 1.046 \\
43 & 300 & 145 & 23.5 & 58.5 & 2498 & 2126 & 1.026 \\
44 & 300 & 138 & 23.6 & 58.3 & 2495 & 2102 & 1.071 \\
45 & 300 & 145 & 25.3 & 62.9 & 2501 & 2125 & 1.011 \\
46 & 300 & 155 & 24.2 & 59.4 & 2479 & 2122 & 1.025 \\
47 & 300 & 135 & 23.2 & 57.7 & 2495 & 2170 & 1.010 \\
48 & 300 & 130 & 22.7 & 56.9 & 2525 & 2150 & 1.019 \\
49 & 300 & 135 & 24.1 & 60.1 & 2509 & 2125 & 1.062 \\
50 & 300 & 150 & 28.3 & 68.0 & 2401 & 2268 & 1.024 \\
51 & 298 & 130 & 28.6 & 68.5 & 2395 & 2324 & 1.014 \\
\midrule
Mean (10) & -- & 141 & 24.6 & 60.7 & 2480 & 2162 & 1.031 \\
\quad s.d. & -- & 8 & 2.2 & 4.3 & 45 & 75 & 0.021 \\
\bottomrule
\end{tabular}
\caption{On-device cost of the single champion (global-best genome),
per seed. Each run streams 300 generations to the Raspberry Pi 4B,
which evaluates the genome on that generation's 50-stock test window.
Times and energies are per pass, averaged over the run; energy is
whole-board energy, and idle power is metered at the start of each run
(Section~\ref{sec:deployment_results}).}
\label{tab:pi-per-seed}
\end{table}

\begin{table}[t]
\centering
\footnotesize
\setlength{\tabcolsep}{2pt}
\begin{tabular}{@{}lrrrr@{}}
\toprule
Generations & Weights & Infer. (ms) & Energy (mJ) & Power (mW) \\
\midrule
0--74 & 57 & 15.2 & 37.1 & 2453 \\
75--149 & 111 & 23.8 & 58.7 & 2484 \\
150--224 & 135 & 28.0 & 69.2 & 2495 \\
225--299 & 152 & 31.5 & 77.8 & 2490 \\
\midrule
Growth & 2.7$\times$ & 2.1$\times$ & 2.1$\times$ & 1.02$\times$ \\
\bottomrule
\end{tabular}
\caption{Single-champion cost against search progress, pooled over
the ten runs of Table~\ref{tab:pi-per-seed}. Each row averages one
quarter of the campaign. Inference time and energy grow with the
evolved network; board power is flat, so the energy growth is time,
not draw (Section~\ref{sec:deployment_results}).}
\label{tab:pi-progression}
\end{table}

\begin{table}[t]
\centering
\footnotesize
\setlength{\tabcolsep}{3pt}
\begin{tabular}{@{}lrrrrrr@{}}
\toprule
Seed & Gen. & Weights & Infer. & Energy & MSE & MSE \\
 & & & (ms) & (mJ) & (ens.) & (memb.) \\
\midrule
45 & 268 & 4,528 & 527 & 1,589 (371) & 0.975 & 1.018 \\
46 & 300 & 4,317 & 444 & 1,378 (371) & 0.975 & 1.011 \\
47 & 300 & 4,584 & 618 & 1,796 (411) & 0.975 & 1.019 \\
48 & 300 & 4,521 & 634 & 1,845 (223) & 0.975 & 1.020 \\
49 & 300 & 4,651 & 602 & 1,746 (213) & 0.975 & 1.016 \\
50 & 243 & 4,328 & 498 & 1,527 (396) & 0.975 & 1.016 \\
\midrule
Mean (6) & -- & 4,488 & 554 & 1,647 (331) & 0.975 & 1.016 \\
\quad s.d. & -- & 137 & 76 & 180 (89) & 0.000 & 0.003 \\
\bottomrule
\end{tabular}
\caption{On-device cost of the island-champion ensemble, per seed. A
pass runs all 40 island champions and averages their predictions, so
weights and costs are totals over those networks; energy is
whole-board energy, with the marginal cost above idle in brackets.
MSE (ens.) is the ensemble's error and MSE (memb.) the mean error of
its members. Seeds that stopped before generation 300 are averaged
over the generations they completed
(Section~\ref{sec:deployment_results}).}
\label{tab:pi-island}
\end{table}

\begin{table}[t]
\centering
\footnotesize
\setlength{\tabcolsep}{2pt}
\begin{tabular}{@{}lrrrrr@{}}
\toprule
Option & Runs & Weights & Infer. & Energy & MSE \\
 & & & (ms) & (mJ) & \\
\midrule
Single champion      & 10 & 114   & 24.6  & 61 (7)      & 1.031 \\
Island-champion ens. & 6  & 4,492 & 556.2 & 1,652 (328) & 0.975 \\
\midrule
Ratio & -- & 39$\times$ & 23$\times$ & 27$\times$ (45$\times$) & 0.946$\times$ \\
\bottomrule
\end{tabular}
\caption{Cost of one pass for the two deployment options, pooled over
all generations of all runs; weights are averaged over the campaign
(the per-quarter values of Table~\ref{tab:pi-progression} average to
114). Energy is whole-board energy, with the marginal cost above idle
in brackets (Section~\ref{sec:deployment_results}).}
\label{tab:pi-deployment}
\end{table}

\subsection{Source of the Ensembling Gain}
\label{app:res-rule}

Table~\ref{tab:diversity} gives the quantities behind the
equicorrelated-averaging check of Section~\ref{sec:res-rule}: for each
width, the mean rank IC $m$ of an individual island champion, the mean
pairwise Spearman correlation $\rho$ between member cross-sections,
the measured ensemble IC, and the prediction $m\sqrt{N/(1+(N-1)\rho)}$,
which is within 3\% of the measurement at every width.
Table~\ref{tab:invariance} recomputes the ensemble-minus-champion
difference inside all sixteen configurations of the tuning screen
(Appendix~\ref{app:onenas}); it is positive in 16 of 16 cells with mean
$+8.3$ points. Those cells run at the 8-island registered width, so
their ensembles have 8 members rather than 40, which is why their
gains are smaller than the $+27.8$ points of the 40-island tuning-span
fleet in Table~\ref{tab:m1}.

\begin{table}[t]
\centering
\footnotesize
\begin{tabular}{@{}l@{\hspace{3pt}}c@{\hspace{3pt}}c@{\hspace{3pt}}c@{\hspace{3pt}}c@{\hspace{3pt}}c@{}}
\toprule
Fleet & $N$ & $\rho$ & Member $m$ & Ensemble & Predicted \\
\midrule
20 islands & 20 & 0.166 & $+0.0068$ & $+0.0145$ & $+0.0149$ \\
40 islands & 40 & 0.150 & $+0.0065$ & $+0.0153$ & $+0.0156$ \\
60 islands & 60 & 0.163 & $+0.0067$ & $+0.0157$ & $+0.0160$ \\
\bottomrule
\end{tabular}
\caption{Ensemble diversity mechanism (Section~\ref{sec:res-rule}):
member rank IC $m$, pairwise member prediction correlation $\rho$,
measured ensemble rank IC, and the equicorrelated averaging prediction
$m\sqrt{N/(1+(N-1)\rho)}$, 2022--2024, Overlapping rule.}
\label{tab:diversity}
\end{table}

\begin{table}[t]
\centering
\footnotesize
\begin{tabular}{@{}lr@{\quad}lr@{}}
\toprule
Configuration & $\Delta$net & Configuration & $\Delta$net \\
\midrule
C0 (registered)      & $+3.9$  & NTS 600            & $+12.8$ \\
Select: IC-gated     & $+0.5$  & Rounds 2           & $+11.0$ \\
Select: IC           & $+2.8$  & PER $\alpha$ 0.4   & $+4.0$  \\
Target: 5-day CS     & $+4.5$  & PER $\alpha$ 0.8   & $+11.7$ \\
5-day CS + IC-gated  & $+1.8$  & PER $\lambda{\times}3$ & $+20.6$ \\
Elite 5 / gen 10     & $+9.6$  & BP epochs 20       & $+12.5$ \\
Elite 5 / gen 15     & $+11.1$ & Repop.\ 25         & $+7.0$  \\
Repop.\ 100          & $+9.6$  & PER off (uniform)  & $+9.3$  \\
\bottomrule
\end{tabular}
\par\smallskip
{\footnotesize Mean $+8.3$; $n{=}6$ runs per cell (2 panels $\times$ 3
seeds), all at the 8-island registered width. Abbreviations as in
Table~\ref{tab:hps}.\par}
\caption{Ensembling gain (ensemble $-$ single champion, within-run, net
\% on 2016--2019, Overlapping rule) inside every trained
hyperparameter configuration of the tuning screen; positive in 16/16
cells (Section~\ref{sec:res-rule}).}
\label{tab:invariance}
\end{table}

\subsection{Trading Rule and Factor Exposure}
\label{app:res-robust}

Table~\ref{tab:books} runs the three rules of Appendix~\ref{app:books}
on the same 40-island predictions and costs, and Table~\ref{tab:alphas}
gives the factor regression of Appendix~\ref{app:metrics} for every
arm of Table~\ref{tab:window}; both are discussed in
Section~\ref{sec:res-robust}, and the cost-sensitivity table of that
section is in the main text (Table~\ref{tab:costs}).

\begin{table}[t]
\centering
\footnotesize
\begin{tabular}{@{}l@{\hspace{5pt}}r@{\hspace{5pt}}r@{\hspace{5pt}}r@{\hspace{5pt}}r@{}}
\toprule
Rule & 2022 & 2023 & 2024 & 2022--24 \\
\midrule
Overlapping     & $+6.1$\se{0.8}  & $+12.4$\se{0.7} & $+6.3$\se{0.5}  & $+27.5$\se{1.1} \\
Banded$^{a}$    & $+12.5$\se{1.2} & $+14.6$\se{1.3} & $+3.4$\se{0.9}  & $+33.9$\se{1.8} \\
Cond.\ L/S$^{b}$ & $+1.3$\se{1.6}  & $+12.8$\se{1.0} & $+10.8$\se{1.2} & $+25.0$\se{1.8} \\
\bottomrule
\end{tabular}
\par\smallskip
{\footnotesize
$^{a}$Enter at rank $\le$10, exit past rank 35.
$^{b}$Algorithm~\ref{alg:long-short} with $k=10$: rebalance only when
the top-10 predictions are all positive and the bottom-10 all
negative; on rank-normal predictions the gate fires daily.\par}
\caption{Trading rule: net return (\%) of the ONE-NAS ensemble (40
islands) under the three rules on the same predictions and costs, mean
over $V1$--$V4$, 10 seeds; the Overlapping row is the 40-island row
of Table~\ref{tab:window} (Section~\ref{sec:res-robust}).}
\label{tab:books}
\end{table}

\begin{table}[t]
\centering
\footnotesize
\begin{tabular}{@{}l@{\hspace{4pt}}r@{\hspace{4pt}}r@{\hspace{4pt}}r@{\hspace{4pt}}r@{}}
\toprule
Arm & bps/day & \%/yr & NW $t$ & Beta \\
\midrule
ONE-NAS ($N{=}60$)        & $+4.43$ & $+11.2$ & $+2.73$ & $+0.120$ \\
ONE-NAS ($N{=}40$)        & $+4.05$ & $+10.2$ & $+2.56$ & $+0.109$ \\
ONE-NAS ($N{=}20$)        & $+3.99$ & $+10.0$ & $+2.77$ & $+0.114$ \\
Periodic LSTM (monthly)    & $+2.39$ & $+6.0$  & $+2.01$ & $+0.046$ \\
Online LSTM                & $+1.93$ & $+4.9$  & $+1.61$ & $+0.069$ \\
Online GRU                 & $+1.75$ & $+4.4$  & $+1.91$ & $+0.033$ \\
Online AR                  & $-0.81$ & $-2.1$  & $-0.64$ & $+0.045$ \\
\midrule
Buy \& hold                & $-0.91$ & $-2.3$  & $-0.72$ & $+0.802$ \\
\bottomrule
\end{tabular}
\caption{Factor-adjusted alphas: pooled daily book returns of each arm
of Table~\ref{tab:window} regressed on the Fama--French three factors
plus momentum, 2022--2024, Overlapping rule, Newey--West $t$ with lag
10 (Section~\ref{sec:res-robust}).}
\label{tab:alphas}
\end{table}